\documentclass{llncs}

\usepackage{cmap}
\usepackage[utf8]{inputenc}
\usepackage{latexsym}
\usepackage[english]{babel}
\usepackage{graphicx,color}
\usepackage{hyperref}
\usepackage[all]{hypcap}
\usepackage{enumerate}
\usepackage{booktabs}
\usepackage{cite}
\usepackage{newunicodechar}
\usepackage{listings}

\newunicodechar{×}{\ensuremath\times}
\newunicodechar{→}{\ensuremath\rightarrow}

\begin{document}
\title{Generation of Zoomable maps\\ with Rivers and Fjords}
\subtitle{Unpublished 2022}

\author{Torben {\AE}. Mogensen \and Emil N. Isenbecker}

\institute{Department of Computer Science, University of Copenhagen\\ \email{torbenm@di.ku.dk} \and Department of Computer Science, University of Copenhagen\\ \email{lwr500@alumni.ku.dk}}


%
%

\maketitle


\begin{abstract}
This paper presents a method for generating maps with rivers
and fjords. The method is based on recursive subdivision of triangles
and allows unlimited zoom on details without requiring generation of a
full map at high resolution.
\end{abstract}

\section{Introduction}

Procedurally generated maps are often used in computer games, where
they allow variability when playing a game multiple times.  For
strategy games, a full map of the entire playing area is typically
generated, but for first-person-view games, storing a full map with
all details can be unrealistic, so details are typically generated on
the fly.  This requires \emph{zoomable} maps, where a high-resolution
local map can be generated without having to generate the full map at
this resolution.  You may even want to generate different levels of
detail for the near and far parts of the map.  A similar requirement
exists for pen-and-paper role-playing games, where you might want both
a world map and local maps for individual islands and continents.

Many algorithms exist for generating topographical maps, basically
generating altitude information for every point and then colouring the
points according to some colour map.  Examples include the
diamond-square algorithm~\cite{FFC:82}, the square-square
algorithm~\cite{Miller:86a}, Fourier-transform-based
methods~\cite{Bourke:fft}, great circle
displacement~\cite{Voss:85,Bourke:fake,Olsson}, simulated plate
tectonics~\cite{Burke:plate}, and tetrahedral
subdivision~\cite{Mogensen:2009}.  Most of these suffer from lack of
zoomability: Generating a high resolution image of a small section of
a map requires almost as much computation time as generating a
high-resolution image of the full map.  Exceptions are methods based
on recursive subdivision such as diamond-square and tetrahedral
subdivision.

None of these methods natually support generation of rivers.  For any
fixed-size map, rivers can be placed after the map is generated, but
this prevents zooming, as rivers can not be consistently placed on a
local map without knowing their placement on the global map -- we
need, at least, to know where rivers flow into the selected map
fragment.  So the only zoom option is to pregenerate the full map at
the resolution of the highest zoom and then show only the selected
part.  But that is very costly.  Ideally, a zoomed map should be
generated at a cost which is close to proportional to the size (in
pixels) of the shown portion of the map and not of the full map.

Generation of realistic river networks using Strahler-Horton analysis
has been proposed~\cite{Viennot:1990}, but these do not follow any
terrain.  The challenge is to co-generate terrain and rivers while
being zoomable.

\begin{figure}

\begin{tabular}{cc}
\includegraphics[width=0.48\textwidth]{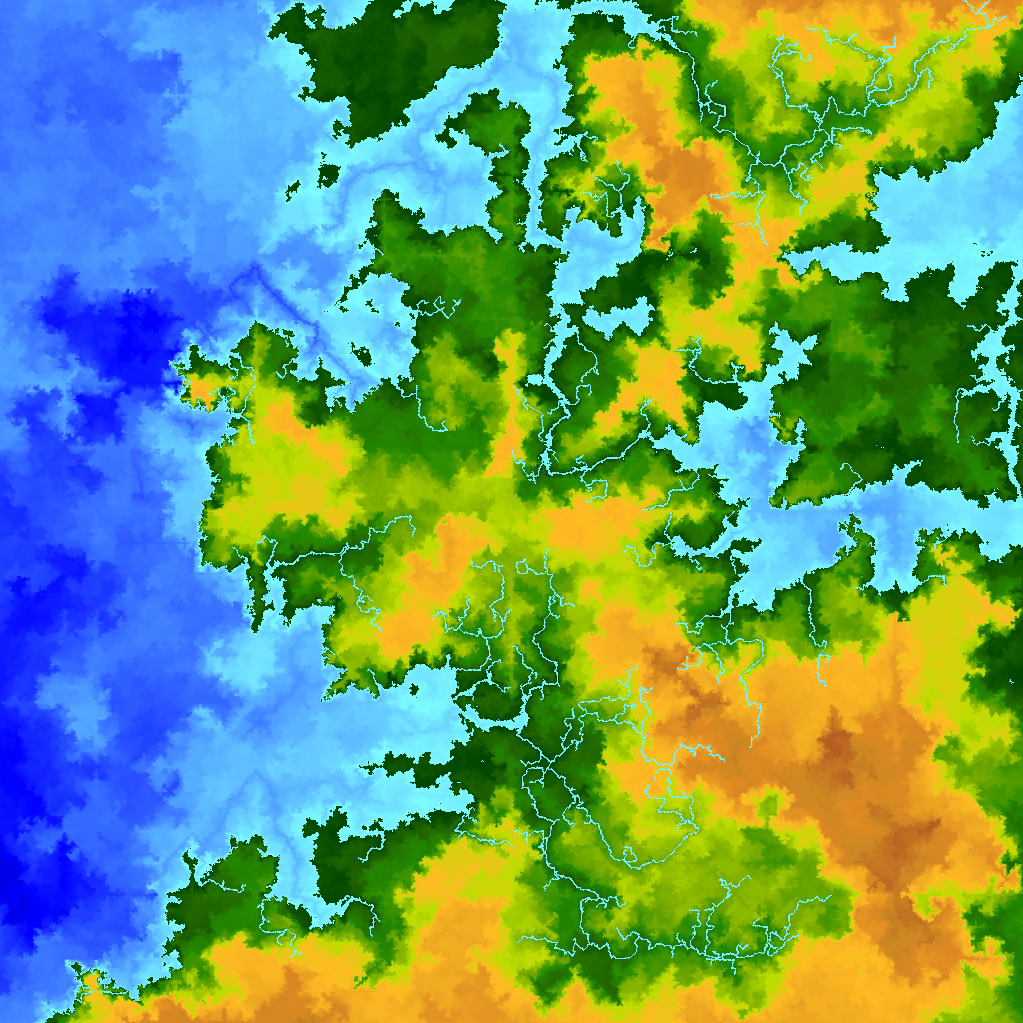} & 

\includegraphics[width=0.48\textwidth]{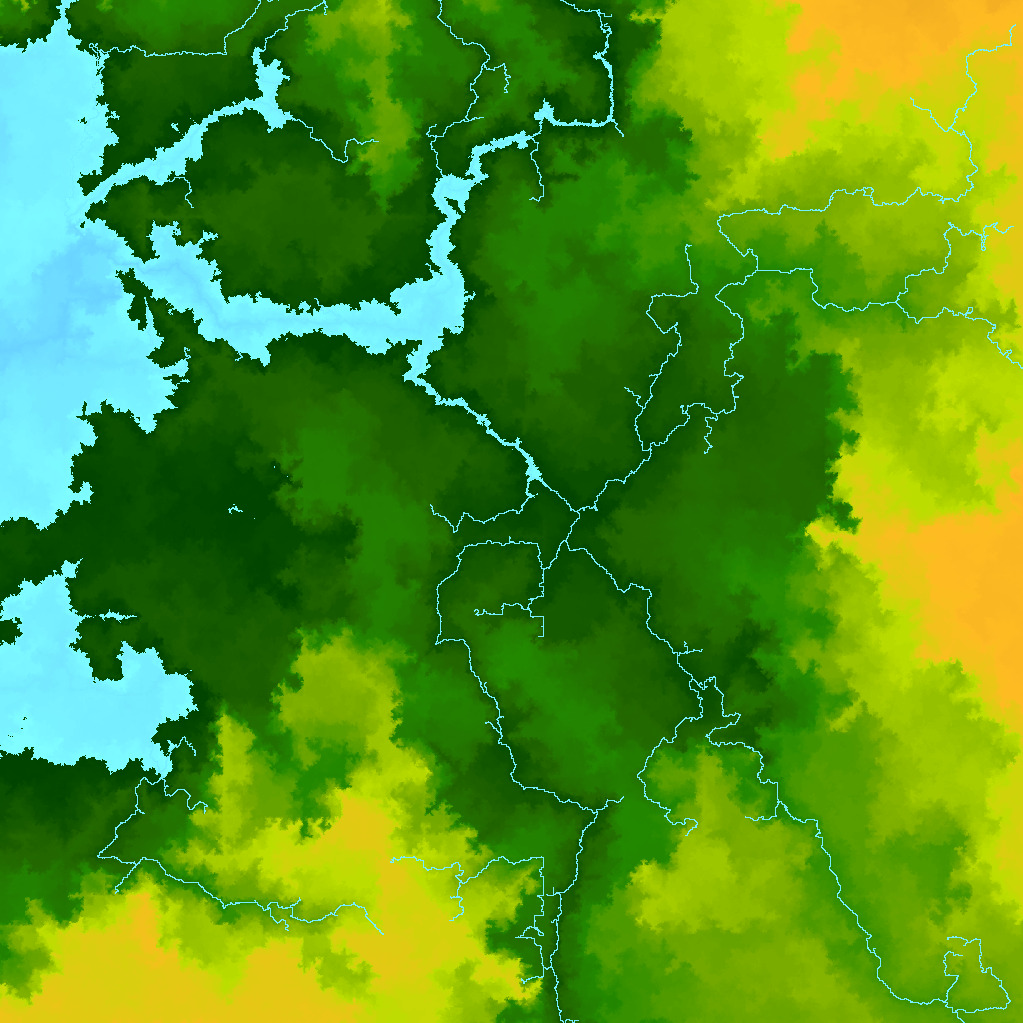} \\[1ex]

\includegraphics[width=0.48\textwidth]{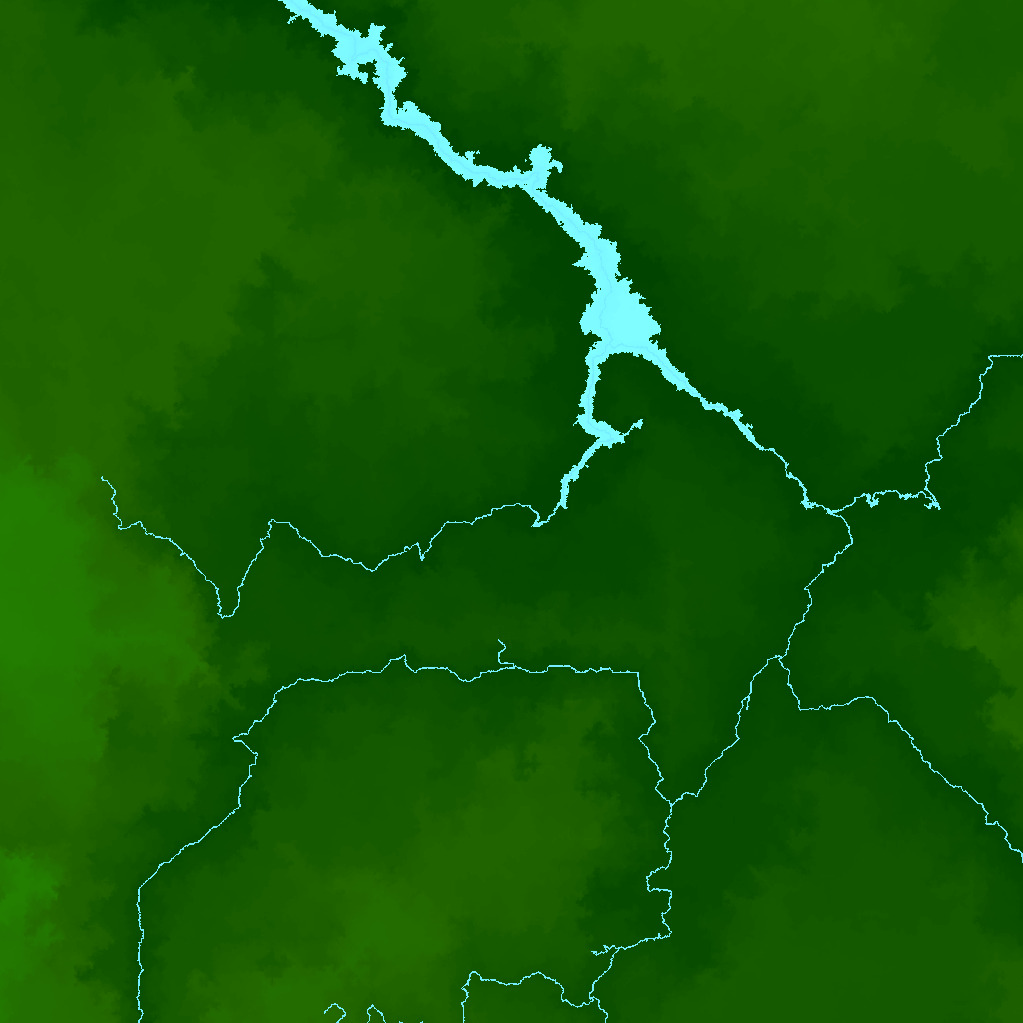} & 

\includegraphics[width=0.48\textwidth]{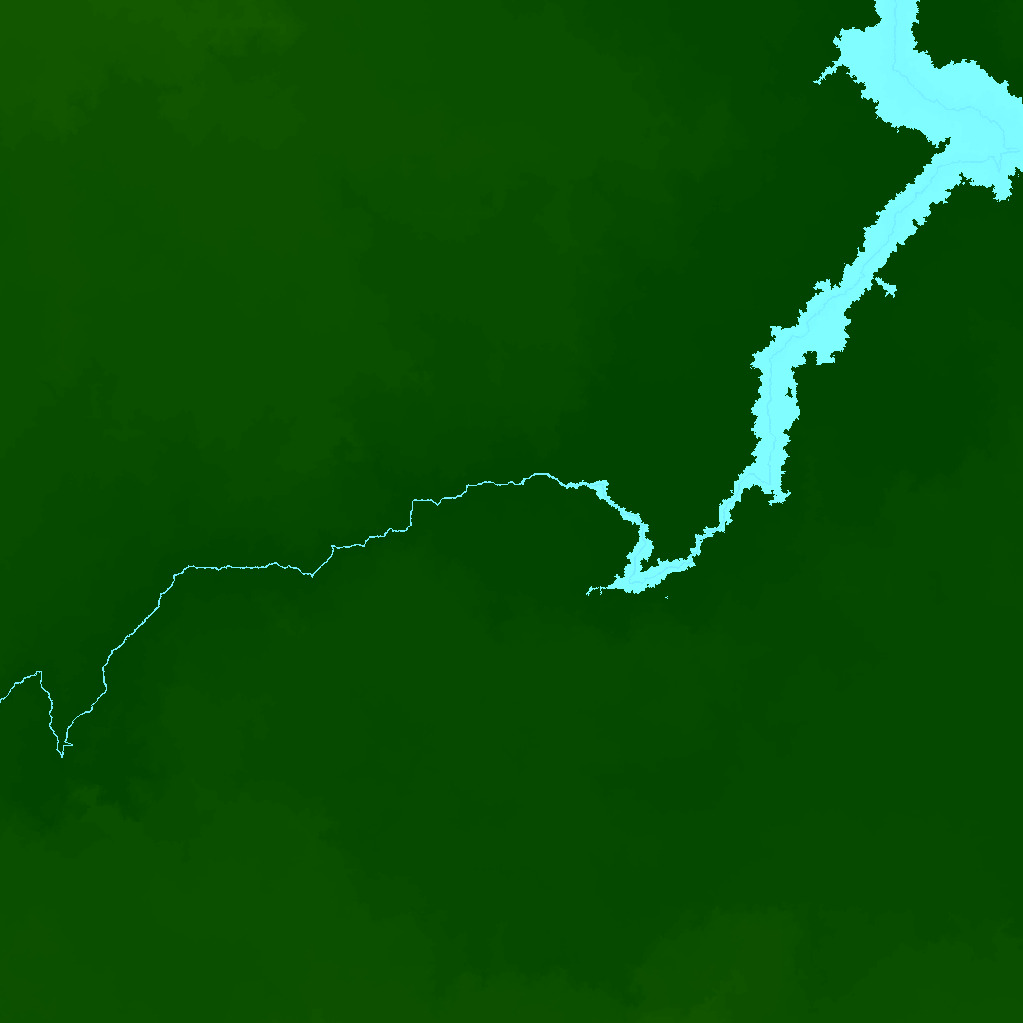}
\end{tabular}
\caption{Map showing rivers and fjords at zoom 1, 5, 25 and 125}\label{riversmap}
\end{figure}

This paper shows a method for doing just that.  The method is based on
triangle subdivision, so it suffers from the same kind of artefacts as
the diamond-square algorithm, but it does generate zoomable maps with
decent-looking river networks and fjords, which to the authors'
knowledge has not been done previously.  Figure~\ref{riversmap} shows
an example of such a map at four different degrees of zoom.  On the
top-left picture it is clear that the river network actually continues
as trenches in the fjord.


\section{Map Generation by Triangle Subdivision}

The first step is to present a simple map generation method using
triangular subdivision.  We use isosceles, right-angled triangles as
shown in Figure~\ref{triangle}(a).  We divide this into two half-size
triangle by drawing a new edge $e_3$ from the vertex $V_0$ to the middle of the
opposite edge $e_0$, splitting this into two new edges $e_4$ and
$e_5$, as shown in Figure~\ref{triangle}(b).  The new vertex at the
middle of $e_0$ is called $V_3$.

\begin{figure}
\setlength{\unitlength}{2ex}

\begin{picture}(20,20)
\put(2,18){\line(1,-1){16}}
\put(2,2){\line(1,0){16}}
\put(2,2){\line(0,1){16}}
\put(1,1){$V_0$}
\put(1,19){$V_1$}
\put(19,1){$V_2$}
\put(10.2,10.2){$e_0$}
\put(0.5,9){$e_2$}
\put(9,0.5){$e_1$}
\end{picture}

\begin{center}
(a) Isosceles, right-angled triangle.
\end{center}

\bigskip

\begin{picture}(20,20)
\put(2,18){\line(1,-1){16}}
\put(2,2){\line(1,0){16}}
\put(2,2){\line(0,1){16}}
\put(1,1){$V_0$}
\put(1,19){$V_1$}

\put(19,1){$V_2$}
\put(0.5,9){$e_2$}
\put(9,0.5){$e_1$}

\put(2,2){\line(1,1){8}}
\put(10.2,10.2){$V_3$}
\put(14.2,6.2){$e_5$}
\put(10.2,10.2){$V_3$}
\put(6.3,5.7){$e_3$}
\put(6.2,14.2){$e_4$}
\end{picture}

\begin{center}
(b) Subdivision of triangle.
\end{center}

\caption{Triangle Subdivision}\label{triangle}
\end{figure}

Each vertex is represented by a record that contains the coordinates
$(x,y)$ of the vertex (each ranging from 0.0 to 1.0), the altitude $h$
at the vertex (ranging from -1.0 to 1.0), and a pseudorandom value $s$
in the range -1.0 to 1.0.  We generate the attributes of $V_3$ by the
formulae:

\[
\begin{array}{lcl}
V_3.x &=& (V_1.x+V_2.x)/2\\
V_3.y &=& (V_1.y+V_2.y)/2\\
V_3.s &=& \mu(V_1.s,\,V_2.s)\\
\delta &=& k_1|V_1-V_2|+k_2|V_1.h-V_2.h|\\
V_3.h &=& (V_1.h+V_2.h)/2 + \delta V_3.s
\end{array}
\]

\noindent
where $\mu$ is a mixing function that takes two pseudorandom values
and produces a new pseudorandom value.  It must be the case that
$\mu(s_1,s_2) = \mu(s_2,s_1)$ for all $s_1$ and $s_2$.  We will also
use $\nu(s) = \mu(s,s)$.

$\delta$ is a measure of how much the altitude can differ from the
average of $V_1.h$ and $V_2.h$.  $k_1$ and $k_2$ are constants that
determine how much the altitude varies as a function of the distance
between the vertices ($|V_1-V_2|$) and the altitude difference between
the vertices ($|V_1.h-V_2.h|$), respectively. For the examples used in
this paper, we use $k_1=0.32$ and $k_2=0.55$.  It is possible (though
unlikely) that this gives a $V_3.h$ outside the range -1.0 to 1.0.  If
this happens, we cap it to be inside the range.

Note that $V_0.h$ is not used when finding $V_3.h$.  This is because
the edge $e_0$ can be shared with another triangle that does not have
access to $V_0$.  This is the main difference between this method and
the diamond-square method: Only two altitudes are averaged when
finding the altitude of a new point, where the diamond-square method
averages four altitudes when finding the altitude of a new point.

\begin{figure}

\includegraphics[width=0.5\textwidth]{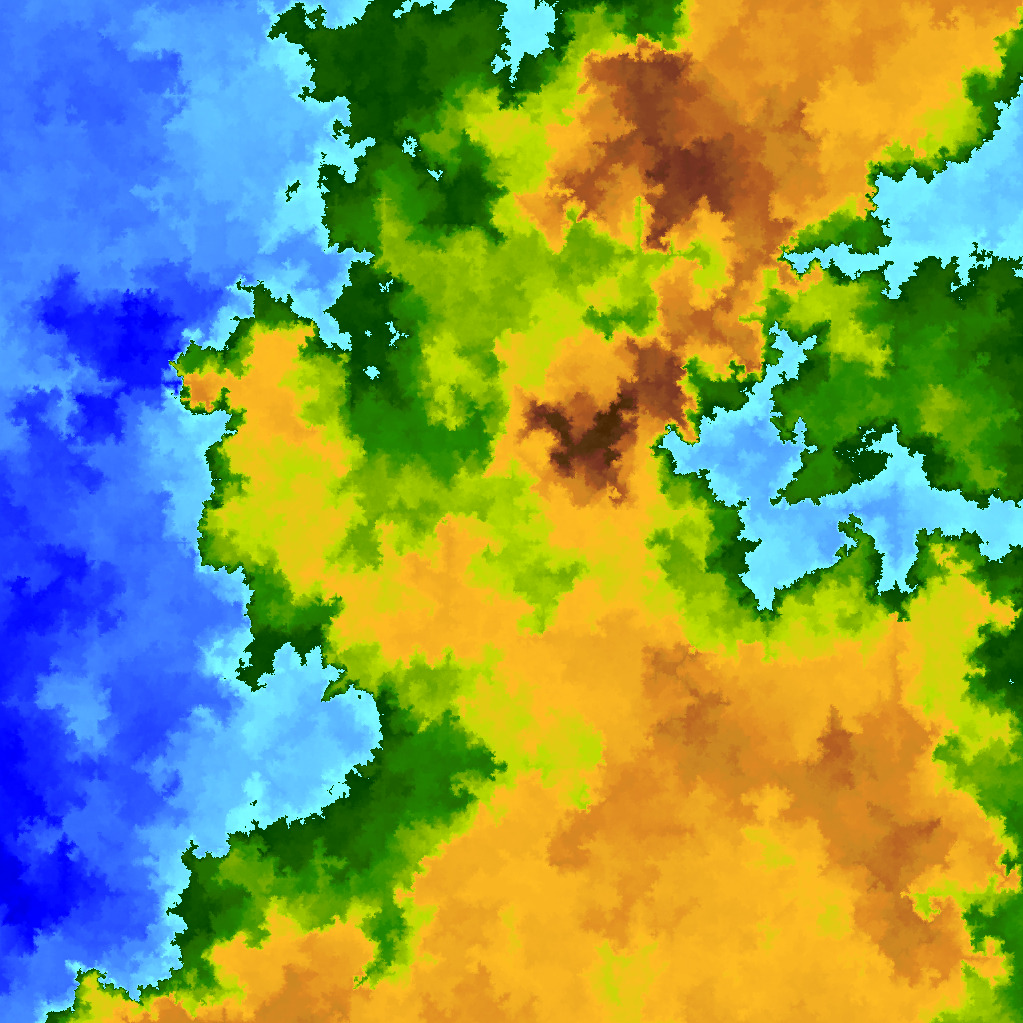}

\caption{Map generated by triangle subdivision}\label{norivers}
\end{figure}

We continue recursively subdividing the triangle into ever-smaller
triangles, until a triangle is the size of one pixel.  At this point,
we use $V_3.h$ to colour the corresponding point on the map.  The maps
shown in this paper uses blue shades for negative altitudes and green,
brown, and white shades for positive altitudes.  A map produced by
this simple algorithm is shown in Figure~\ref{norivers}.

\section{Adding Rivers}

We will now extend the simple map generator to include rivers.  We
will do so using the edges of the triangle: An edge $e_i$ has an
\emph{optional} altitude attribute $e_i.h$.  If this attribute is
present, it indicates that \emph{somewhere} on edge $e_i$ a river
crosses at altitude $e_i.h$.  If the attribute is absent, no river
flows through the edge.  When subdividing a triangle, we use the
attributes on $e_0$, $e_1$ and $e_2$ together with the attributes on
$V_0$, $V_1$ and $V_2$ to generate the attributes for $e_3$, $e_4$,
and $e_5$.  Moreover, we allow the altitude attribute $e_0.h$ to
influence $V_3.h$.  Basically, we want that, when the subdivision
finishes, $V_3.h$ denotes the altitude of the corresponding point of
the river (if any).

\subsection{The long edge}

We recall that the attributes of $V_3$, $e_4$ and $e_5$ can only
depend on the attributes of $V_1$, $V_2$ and $e_0$, since the edge
$e_0$ can be shared with another triangle that does not have access to
$V_0$, $e_1$ and $e_2$.  So let us start with $V_3$, $e_4$ and $e_5$
and then handle $e_3$ later.

\begin{figure}
  \begin{center}
    \setlength{\unitlength}{1ex}
    \begin{picture}(30,20)
      \put(0,5){\line(1,0){10}}
      \put(0,5){\line(0,1){10}}
      \put(10,5){\line(-1,1){10}}
      \put(5,10){\makebox(0,0){\rotatebox{45}{$\sim$}}}

      \put(12,11){\vector(2,1){5}}
      \put(15,10){\makebox(0,0){?}}
      \put(12,9){\vector(2,-1){5}}

      \put(20,12){\line(1,0){10}}
      \put(20,12){\line(0,1){10}}
      \put(30,12){\line(-1,1){10}}
      \put(20,12){\line(1,1){5}}
      \put(22.5,19.5){\makebox(0,0){\rotatebox{45}{$\sim$}}}

      \put(20,-2){\line(1,0){10}}
      \put(20,-2){\line(0,1){10}}
      \put(30,-2){\line(-1,1){10}}
      \put(20,-2){\line(1,1){5}}
      \put(27.5,0.5){\makebox(0,0){\rotatebox{45}{$\sim$}}}
    \end{picture}
  \end{center}
\caption{When there is a river on the long edge}\label{long-edge-river}
\end{figure}
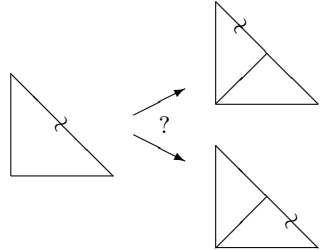

We noted that $e_0$ has an attribute $e_0.h$ if and only if a river
crosses $e_0$.  If this is the case, we have to choose whether this
river crosses the half-edge $e_4$ or the half-edge $e_5$, as shown in
Figure~\ref{long-edge-river}.  We do not allow a river to cross both
$e_4$ and $e_5$, as that could potentially lead to circular river
segments that do not connect to sea.  We use a simple method for this:
If $e_0.h$ is closer to $V_1.h$ than to $V_2.h$, the river flows
through $e_4$, otherwise it flows through $e_5$. The attribute of
$e_0$ is copied to the chosen half-edge, and the other half-edge has
no attribute.  Ties can be broken by looking at the other attributes
of $V_1$ and $V_2$, but ties are extremely rare when floating-point
numbers are used.  If a river flows through $e_4$, $e_4.h$ is used
instead of $V_1.h$ when determining $V_3.h$, and if a river flows
through $e_5$, $e_4.h$ is used instead of $V2.h$.  If no river flows
through $e_0$, we determine $V_3.h$ as before.  We can describe the
determination of $e_4$, $e_5$, and $V_3.h$ with the following
pseudocode:

\[\begin{array}{l}
\texttt{if $e_0$ has an attribute $e_0.h$, then}\\
\hspace{3ex}\texttt{if $|e_0.h-V_1.h| < |e_0.h-V_2.h|$ then}\\
\hspace{6ex}e_4.h = e_0.h;\\
\hspace{6ex}e_5.h = \texttt{Absent};\\
\hspace{6ex}V_3.h = (e_4.h+V_2.h)/2 + \delta V_3.s\\
\hspace{3ex}\texttt{else}\\
\hspace{6ex}e_5.h = e_0.h;\\
\hspace{6ex}e_4.h = \texttt{Absent};\\
\hspace{6ex}V_3.h = (e_5.h+V_1.h)/2 + \delta V_3.s\\
\texttt{else}\\
\hspace{3ex}e_5.h = \texttt{Absent};\\
\hspace{3ex}e_4.h = \texttt{Absent};\\
\hspace{3ex}V_3.h = (V_1.h+V_2.h)/2 + \delta V_3.s
\end{array}\]

\noindent
The other attributes of $V_3$ are calculated as before.

\subsection{The short edge $e_3$}

The complicated part is determining the attribute (if any) of $e_3$.
Note that this can (and does) depend on all vertices and all edges.

\begin{figure}
  \begin{center}
    \setlength{\unitlength}{1ex}
    \begin{picture}(30,20)
      \put(0,5){\line(1,0){10}}
      \put(0,5){\line(0,1){10}}
      \put(10,5){\line(-1,1){10}}

      \put(12,11){\vector(2,1){5}}
      \put(15,10){\makebox(0,0){?}}
      \put(12,9){\vector(2,-1){5}}

      \put(20,12){\line(1,0){10}}
      \put(20,12){\line(0,1){10}}
      \put(30,12){\line(-1,1){10}}
      \put(20,12){\line(1,1){5}}

      \put(20,-2){\line(1,0){10}}
      \put(20,-2){\line(0,1){10}}
      \put(30,-2){\line(-1,1){10}}
      \put(20,-2){\line(1,1){5}}
      \put(22.5,0.5){\makebox(0,0){\rotatebox{-45}{$\sim$}}}
    \end{picture}
  \end{center}
\caption{When there is no river on any outer edge}\label{no-river}
\end{figure}
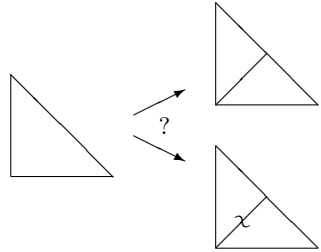

\subsubsection{No surrounding edge has a river}

We start with the case where there is no river on $e_1$, $e_2$, $e_4$
or $e_5$.  We then need to determine if a river should flow through
$e_3$ or not, as illustrated in Figure~\ref{no-river}.  If $V_1$ is
above water and $V_2$ is below water, there is a possibility of a
river flowing from somewhere in the triangle $V_1,\,V_0,\,V_3$ to
somewhere in the triangle $V_2,\,V_0,\,V_3$, but only if there is no
closer exit.  So we rule that both $V_0$ and $V_3$ should be higher
than $V_2$.  We also prefer rivers with some steepness over very flat
rivers, so we want $V_1$ to be higher above water than $V_2$ if they
are far apart than if they are close.  Additionally, we want the
height of the river at $e_3$ to be between $V_2.h$ and the minimum of
$V_0.h$ and $V_3.h$.  In pseudocode, this is

\[\begin{array}{l}
\texttt{if $V_1.h>k_3 \wedge V_2.h < k_4 \wedge V_2.h <
  V_0.h \wedge V_2.h < V_3.h$}\\
\texttt{then}\\
\hspace{3ex}e_3.h = \beta(V_2.h,\,\min(V_0.h,\,V_3.h),\,\mu(V_0.s,\,V_3.s));\\
\texttt{else}\\
\hspace{3ex}e_3.h = \texttt{Absent};\\
\end{array}\]

\noindent
Where $k_3$ and $k_4$ are the minimal above-land and below-sea
altitudes for $V_1.h$ and $V_2.h$.  In the examples in this paper, we
have use $k_3=0.1$ and $k_4=-0.1$. $\beta(h_1,h_2,s)$ finds an
altitude between $h_1$ and $h_2$ using the pseudorandom value
$s$. This should tend towards the middle of the interval, so (since
$-1\leq s \leq 1$), we define $\beta(x,y,s) = (x+y+s^3(x-y))/2$. Other
definitions can be used as long as the result tends towards the
average of $x$ and $y$.

We have in the examples shown here chosen that there is always a river
if the preconditions are met, but you can add a pseudorandom element
if you want fewer rivers.

The case where $V_1$ is below water and $V_2$ is above water is
handled symmetrically.

\subsubsection{One surrounding edge has a river}

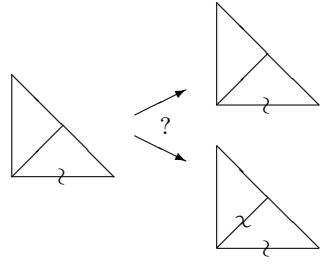
\begin{figure}
  \begin{center}
    \setlength{\unitlength}{1ex}
    \begin{picture}(30,20)
      \put(0,5){\line(1,0){10}}
      \put(0,5){\line(0,1){10}}
      \put(0,5){\line(1,1){5}}
      \put(10,5){\line(-1,1){10}}
      \put(5,5){\makebox(0,0){\rotatebox{90}{$\sim$}}}

      \put(12,11){\vector(2,1){5}}
      \put(15,10){\makebox(0,0){?}}
      \put(12,9){\vector(2,-1){5}}

      \put(20,12){\line(1,0){10}}
      \put(20,12){\line(0,1){10}}
      \put(30,12){\line(-1,1){10}}
      \put(20,12){\line(1,1){5}}
      \put(25,12){\makebox(0,0){\rotatebox{90}{$\sim$}}}

      \put(20,-2){\line(1,0){10}}
      \put(20,-2){\line(0,1){10}}
      \put(30,-2){\line(-1,1){10}}
      \put(20,-2){\line(1,1){5}}
      \put(25,-2){\makebox(0,0){\rotatebox{90}{$\sim$}}}
      \put(22.5,0.5){\makebox(0,0){\rotatebox{-45}{$\sim$}}}
    \end{picture}
  \end{center}
\caption{When there is a river on one outer edge}\label{one-river}
\end{figure}

If only one of the edges $e_1$, $e_2$, $e_4$ and $e_5$ is crossed by a
river, we want to see if we should extend this through $e_3$.  We show
the case if a river passing through $e_1$ (s shown in
Figure~\ref{one-river}, as the other cases are symmetric.

If there is a river through $e_1$, $V_1.h<0$, and $V_1.h<e_1.h$, the river
might flow down towards $V_1$, but only if there is not a shorter
path to the sea near $V_3$.  So if $V_1.h<0$, $V_1.h<e_1.h$ and $V_3.h>0$, we
add a river to $e_3$ with altitude between that of $e_1$ and $V_1$,
so, $e_3.h = \beta(e_1.h,\,V_1.h,\,\mu(V_3.s,V_0.s))$.

If, on the other hand, $V_1.h>e_1.h$, there is a possibility that the
river flows from somewhere in the triangle $V_1$, $V_0$, $V_3$ towards
$e_1$, but only if all three vertices are higher up than $e_3$. Also,
we might not want to always extend the river all the way to the top of
a mountain, so we add a chance that it will not extend across $e_3$.

I pseudocode, we handle the case of a river only through $e_1$ by

\[\begin{array}{l}
\texttt{if $V_1.h<0 \wedge V_1.h<e1.h \wedge V_3.h>0$ then}\\
\hspace{3ex}e_3.h = \beta(V_1.h,\,e_1.h,\,\mu(V_0.s,\,V_3.s));\\
\texttt{else if $V_1.h> e_1.h \wedge V_0.h>e1.h \wedge V_3.h>e_1.h$}\\
\hspace{3ex}\texttt{then if $|\mu(V_0.s,\,V_3.s)| < k_5$ then}\\
\hspace{6ex}e_3.h = \beta(e_1.h,\,\min(V_1.h,\,V_0.h,\,V_3.h,\,\nu(V_1.s));\\
\hspace{3ex}\texttt{else}\\
\hspace{6ex}e_3.h = \texttt{Absent};\\
\texttt{else}\\
\hspace{3ex}e_3.h = \texttt{Absent};\\
\end{array}\]

\noindent
Where $k_5$ is the probability that we extend a river upwards if the
preconditions are met.  In the examples used in this paper, $k_5=0.7$.

There are symmetric cases for rivers going through $e_2$, $e_4$, or
$e_5$. Where we use $V_3$, above, we use $V_0$ when extending rivers
through $e_4$ or $e_5$.  I.e., we do not use a vertex on the same edge
as the river.

\subsubsection{Two surrounding edges have rivers}

\begin{figure}
  \begin{center}
    \setlength{\unitlength}{1ex}
    \begin{picture}(40,10)
      \put(0,0){\line(1,0){10}}
      \put(0,0){\line(0,1){10}}
      \put(0,0){\line(1,1){5}}
      \put(10,0){\line(-1,1){10}}
      \put(5,0){\makebox(0,0){\rotatebox{90}{$\sim$}}}
      \put(0,5){\makebox(0,0){\rotatebox{0}{$\sim$}}}
      \put(2.5,2.5){\makebox(0,0){\rotatebox{-45}{$\sim$}}}

      \put(15,0){\line(1,0){10}}
      \put(15,0){\line(0,1){10}}
      \put(15,0){\line(1,1){5}}
      \put(25,0){\line(-1,1){10}}
      \put(20,0){\makebox(0,0){\rotatebox{90}{$\sim$}}}
      \put(17.5,7.5){\makebox(0,0){\rotatebox{45}{$\sim$}}}
      \put(17.5,2.5){\makebox(0,0){\rotatebox{-45}{$\sim$}}}

      \put(30,0){\line(1,0){10}}
      \put(30,0){\line(0,1){10}}
      \put(30,0){\line(1,1){5}}
      \put(40,0){\line(-1,1){10}}
      \put(37.5,2.5){\makebox(0,0){\rotatebox{45}{$\sim$}}}
      \put(30,5){\makebox(0,0){\rotatebox{0}{$\sim$}}}
      \put(32.5,2.5){\makebox(0,0){\rotatebox{-45}{$\sim$}}}
    \end{picture}
  \end{center}
\caption{When there are rivers on two opposing outer edges}\label{river-must-cross}
\end{figure}
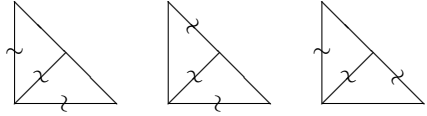

If the two edges that have rivers are on opposite sides of $e_3$, the
case is simple: There \emph{will} be a river on $e_3$ and its altitude
is between that of the two other river edges.  There are three such
cases: rivers on $e_1$ and $e_2$, rivers on $e_1$ and $e_4$, and
rivers on $e_2$ and $e_5$ (since there can not be rivers on both $e_4$
and $e_5$).  We illustrate these cases in
Figure~\ref{river-must-cross}.

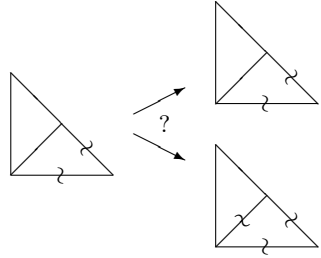
\begin{figure}
  \begin{center}
    \setlength{\unitlength}{1ex}
    \begin{picture}(30,20)
      \put(0,5){\line(1,0){10}}
      \put(0,5){\line(0,1){10}}
      \put(0,5){\line(1,1){5}}
      \put(10,5){\line(-1,1){10}}
      \put(5,5){\makebox(0,0){\rotatebox{90}{$\sim$}}}
      \put(7.5,7.5){\makebox(0,0){\rotatebox{45}{$\sim$}}}

      \put(12,11){\vector(2,1){5}}
      \put(15,10){\makebox(0,0){?}}
      \put(12,9){\vector(2,-1){5}}

      \put(20,12){\line(1,0){10}}
      \put(20,12){\line(0,1){10}}
      \put(30,12){\line(-1,1){10}}
      \put(20,12){\line(1,1){5}}
      \put(25,12){\makebox(0,0){\rotatebox{90}{$\sim$}}}
      \put(27.5,14.5){\makebox(0,0){\rotatebox{45}{$\sim$}}}

      \put(20,-2){\line(1,0){10}}
      \put(20,-2){\line(0,1){10}}
      \put(30,-2){\line(-1,1){10}}
      \put(20,-2){\line(1,1){5}}
      \put(25,-2){\makebox(0,0){\rotatebox{90}{$\sim$}}}
      \put(27.5,0.5){\makebox(0,0){\rotatebox{45}{$\sim$}}}
      \put(22.5,0.5){\makebox(0,0){\rotatebox{-45}{$\sim$}}}
    \end{picture}
  \end{center}
\caption{Potentially branching river}\label{possible-branch}
\end{figure}

The case where the two river-carrying edges are on the same side of
$e_3$ is more interesting.  It is clear that a river flows from one
edge to the other without crossing $e_3$, but there is a possibility
that the river branches with a branch entering through $e_3$, as
illustrated in figure~\ref{possible-branch}.  This branch must be
higher up than the lowest of the river-carrying outside edges, as we
do not allow a river to split downwards.\footnote{This can happen in
  river deltas, but we do not handle this case here.}  We want all
three vertices of the triangle on the opposite edge of $e_3$ to be
higher than the lowest of the river-carrying outside edges, and the
river on $e_4$ must have an altitude between the lowest of the three
vertices and the lowest of the river-carrying outside edges.
Additionally, we might not always want a branch, so we add a
probability.

The case where the edges $e_1$ and $e_5$ carry rivers is shown in
pseudocode below. The case where the edges $e_1$ and $e_5$ carry
rivers is symmetric.

\[\begin{array}{l}
\texttt{if $\min(V_1.h,\,V_0.h,\,V_3.h) > \min(e_1.h,\,e_5.h)$}\\
\texttt{then if $|\mu(V_0.s,\,V_3.s)| < k_6|V_1-V_2|$ then}\\
\hspace{6ex}e_3.h = \beta(\min(V_1.h,\,V_0.h,\,V_3.h),\,\min(e_1.h,\,e_5.h),\,\nu(V_1.s))\\
\hspace{3ex}\texttt{else}\\
\hspace{6ex}e_3.h = \texttt{Absent};\\
\texttt{else}\\
\hspace{3ex}e_3.h = \texttt{Absent};\\
\end{array}\]

\noindent
where $k_6|V_1-V_2|$ is the probability that the river branches if the
preconditions are met.  Note that this means that, on a larger scale,
rivers are more likely to branch than on a smaller scale.  We use
$k_6=2$.

\subsubsection{Three surrounding edges have rivers}

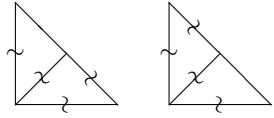
\begin{figure}
  \begin{center}
    \setlength{\unitlength}{1ex}
    \begin{picture}(25,10)
      \put(0,0){\line(1,0){10}}
      \put(0,0){\line(0,1){10}}
      \put(0,0){\line(1,1){5}}
      \put(10,0){\line(-1,1){10}}
      \put(5,0){\makebox(0,0){\rotatebox{90}{$\sim$}}}
      \put(0,5){\makebox(0,0){\rotatebox{0}{$\sim$}}}
      \put(2.5,2.5){\makebox(0,0){\rotatebox{-45}{$\sim$}}}
      \put(7.5,2.5){\makebox(0,0){\rotatebox{45}{$\sim$}}}

      \put(15,0){\line(1,0){10}}
      \put(15,0){\line(0,1){10}}
      \put(15,0){\line(1,1){5}}
      \put(25,0){\line(-1,1){10}}
      \put(20,0){\makebox(0,0){\rotatebox{90}{$\sim$}}}
      \put(15,5){\makebox(0,0){\rotatebox{0}{$\sim$}}}
      \put(17.5,7.5){\makebox(0,0){\rotatebox{45}{$\sim$}}}
      \put(17.5,2.5){\makebox(0,0){\rotatebox{-45}{$\sim$}}}
    \end{picture}
  \end{center}
\caption{When there are rivers on three outer edges, they must connect}\label{three-outer-rivers}
\end{figure}

This case is also relatively simple: There is always a river crossing
$e_3$, as illustrated in Figure~\ref{three-outer-rivers}. The only
nontrivial part is finding the altitude of this river.  The two higher
river edges will flow towards the lowest river edge, so the river on
$e_3$ must be higher than the lowest river edge but lower than the
river(s) on the opposite side of the lowest river.

The case where there are rivers on $e_1$, $e_2$, and $e_4$ is handled
by the pseudocode

\[\begin{array}{l}
e_3.h = \beta(e_1.h,\,\min(e_2.h,e_4.h))
\end{array}\]

\noindent
The case where there are rivers on $e_1$, $e_2$, and $e_5$ is handled
symmetrically.

There are no cases with rivers on both $e_4$ and $e_5$, so we have
handled all possible cases.

Figure~\ref{riversmap} shows a map generated by these rules.  Compare
the top-left image to Figure~\ref{norivers}, which is the same map but
without rivers.  Note how the rivers ``carve'' fjords and river
valleys into the land.  Fjords are generated when the river has an
altitude that is below sea level and ``pulls'' nearby land below sea
as well.

\section{Variant: Islands in Fjords}

\begin{figure}

\begin{tabular}{cc}
  \includegraphics[width=0.45\textwidth]{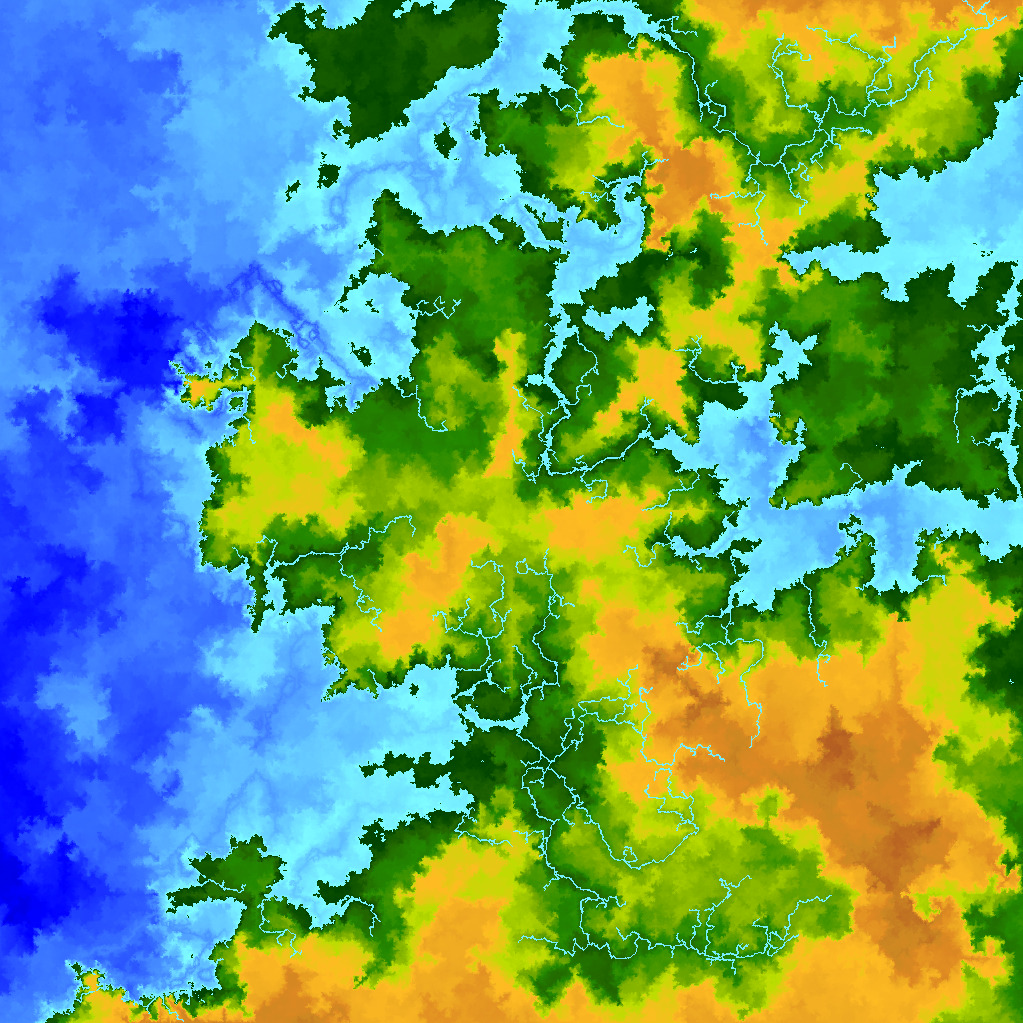}
  & \includegraphics[width=0.45\textwidth]{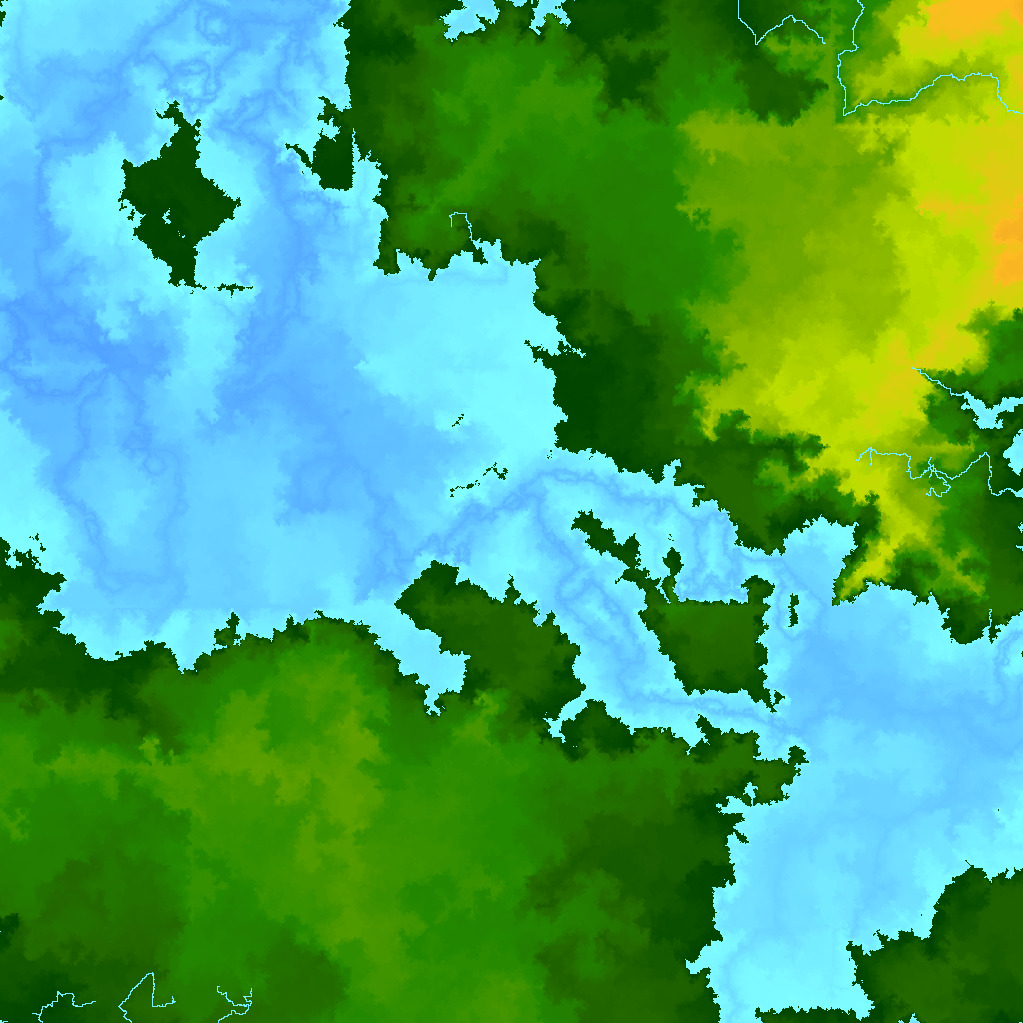}
\end{tabular}
\caption{Islands in fjords}\label{islands}
\end{figure}

If we allow rivers on both halves ($e_4$ and $e_5$) of a split long
edge ($e_0$) of the isosceles triangle, it can potentially lead to
disconnected circular rivers.  But if we do do only if $e_0.h < k_7$,
for some negative $k_7$ (the examples use $k_7 = -0.1$), such
circularity can only happen in fjords.  So instead of circular rivers,
we get islands in fjords and occasional narrow straits crossing land.
Using this variant, we get the maps shown in Figure~\ref{islands}.
The islands are in the fjord near the top of the map.

We have to modify splitting of $e_0$ to

\[\begin{array}{l}
\texttt{if $e_0$ has an attribute $e_0.h$, then}\\
\hspace{3ex}\texttt{if $e_0.h < k_7$ and $|\nu(e_0.s)| < k_8$ then}\\
\hspace{6ex}e_4.h = e_0.h;\\
\hspace{6ex}e_5.h = e_0.h;\\
\hspace{6ex}V_3.h = (e_4.h+e_5.h+V_2.h)/3 + \delta V_3.s\\
\hspace{3ex}\texttt{else if $|e_0.h-V_1.h| < |e_0.h-V_2.h|$ then}\\
\hspace{6ex}e_4.h = e_0.h;\\
\hspace{6ex}e_5.h = \texttt{Absent};\\
\hspace{6ex}V_3.h = (e_4.h+V_2.h)/2 + \delta V_3.s\\
\hspace{3ex}\texttt{else}\\
\hspace{6ex}e_5.h = e_0.h;\\
\hspace{6ex}e_4.h = \texttt{Absent};\\
\hspace{6ex}V_3.h = (e_5.h+V_1.h)/2 + \delta V_3.s\\
\texttt{else}\\
\hspace{3ex}e_5.h = \texttt{Absent};\\
\hspace{3ex}e_4.h = \texttt{Absent};\\
\hspace{3ex}V_3.h = (V_1.h+V_2.h)/2 + \delta V_3.s
\end{array}\]

\noindent
where $k_8$ is a probability of copying the river.  We have used
$k_8=0.15$.  It is not entirely realistic that $e_4.h = e_5.h$, but if
we change one or the other, we can get rivers flowing upward.

We also need to add cases where there are rivers on both $e_4$ and
$e_5$.  If no other edges or all other edges have rivers, we do not
add a river on $e_3$.  If one other edge ($e_1$ or $e_2$) has a river,
we add a river on $e_3$ with an altitude between $e_1.h$ and $e_4.h$
or between $e_2.h$ and $e_5.h$, depending on whether $e_1$ or $e_2$
has a river.

\section{Variant: Penrose Tiles}

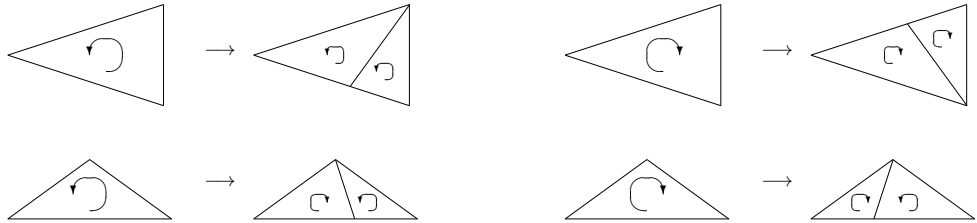
\begin{figure}
\scalebox{0.8}{
  \setlength{\unitlength}{2ex}
  \begin{picture}(60,20)(0,5)
    
    \put(0,15){\rotatebox{18}{\line(1,0){10}}}
    \put(0,15){\rotatebox{-18}{\line(1,0){10}}}
    \put(9.51,11.909){\line(0,1){6.18}}

    \put(6,15){\oval(2,2)[t]}
    \put(6,15){\oval(2,2)[br]}
    \put(5,15){\vector(0,-1){0}}

    \put(12,15){$\longrightarrow$}
    
    \put(15,15){\rotatebox{18}{\line(1,0){10}}}
    \put(15,15){\rotatebox{-18}{\line(1,0){10}}}
    \put(24.51,11.909){\line(0,1){6.18}}
    \put(20.9,18.091){\rotatebox{-36}{\line(0,-1){6.18}}}

    \put(20,15){\oval(1,1)[t]}
    \put(20,15){\oval(1,1)[br]}
    \put(19.5,15){\scalebox{0.75}{\vector(0,-1){0}}}

    \put(23,14){\oval(1,1)[t]}
    \put(23,14){\oval(1,1)[br]}
    \put(22.5,14){\scalebox{0.75}{\vector(0,-1){0}}}

    \put(34,15){\rotatebox{18}{\line(1,0){10}}}
    \put(34,15){\rotatebox{-18}{\line(1,0){10}}}
    \put(43.51,11.909){\line(0,1){6.18}}

    \put(40,15){\oval(2,2)[t]}
    \put(40,15){\oval(2,2)[bl]}
    \put(41,15){\vector(0,-1){0}}

    \put(46,15){$\longrightarrow$}
    
    \put(49,15){\rotatebox{18}{\line(1,0){10}}}
    \put(49,15){\rotatebox{-18}{\line(1,0){10}}}
    \put(58.51,11.909){\line(0,1){6.18}}
    \put(54.9,16.909){\rotatebox{36}{\line(0,-1){6.18}}}

    \put(54,15){\oval(1,1)[t]}
    \put(54,15){\oval(1,1)[bl]}
    \put(54.5,15){\scalebox{0.75}{\vector(0,-1){0}}}

    \put(57,16){\oval(1,1)[t]}
    \put(57,16){\oval(1,1)[bl]}
    \put(57.5,16){\scalebox{0.75}{\vector(0,-1){0}}}

    \put(0,5){\line(1,0){10}}
    \put(0,5){\rotatebox{36}{\line(1,0){6.18}}}
    \put(5,8.618){\rotatebox{-36}{\line(1,0){6.18}}}

    \put(5,6.5){\oval(2,2)[t]}
    \put(5,6.5){\oval(2,2)[br]}
    \put(4,6.5){\vector(0,-1){0}}

    \put(12,7){$\longrightarrow$}

    \put(15,5){\line(1,0){10}}
    \put(15,5){\rotatebox{36}{\line(1,0){6.18}}}
    \put(20,8.618){\rotatebox{-36}{\line(1,0){6.18}}}
    \put(20,8.618){\rotatebox{18}{\line(0,-1){3.8}}}

    \put(19,6){\oval(1,1)[t]}
    \put(19,6){\oval(1,1)[bl]}
    \put(19.5,6){\scalebox{0.75}{\vector(0,-1){0}}}

    \put(22,6){\oval(1,1)[t]}
    \put(22,6){\oval(1,1)[br]}
    \put(21.5,6){\scalebox{0.75}{\vector(0,-1){0}}}

    \put(34,5){\line(1,0){10}}
    \put(34,5){\rotatebox{36}{\line(1,0){6.18}}}
    \put(39,8.618){\rotatebox{-36}{\line(1,0){6.18}}}

    \put(39,6.5){\oval(2,2)[t]}
    \put(39,6.5){\oval(2,2)[bl]}
    \put(40,6.5){\vector(0,-1){0}}

    \put(46,7){$\longrightarrow$}

    \put(49,5){\line(1,0){10}}
    \put(49,5){\rotatebox{36}{\line(1,0){6.18}}}
    \put(54,8.618){\rotatebox{-36}{\line(1,0){6.18}}}
    \put(52.85,8.618){\rotatebox{-18}{\line(0,-1){3.8}}}

    \put(52,6){\oval(1,1)[t]}
    \put(52,6){\oval(1,1)[bl]}
    \put(52.5,6){\scalebox{0.75}{\vector(0,-1){0}}}

    \put(55,6){\oval(1,1)[t]}
    \put(55,6){\oval(1,1)[br]}
    \put(54.5,6){\scalebox{0.75}{\vector(0,-1){0}}}

  \end{picture}
}

\caption{Subdividing Robinson triangles}\label{Robinson}
\end{figure}

\begin{figure}
  \begin{center}
  \includegraphics[width=0.6\textwidth]{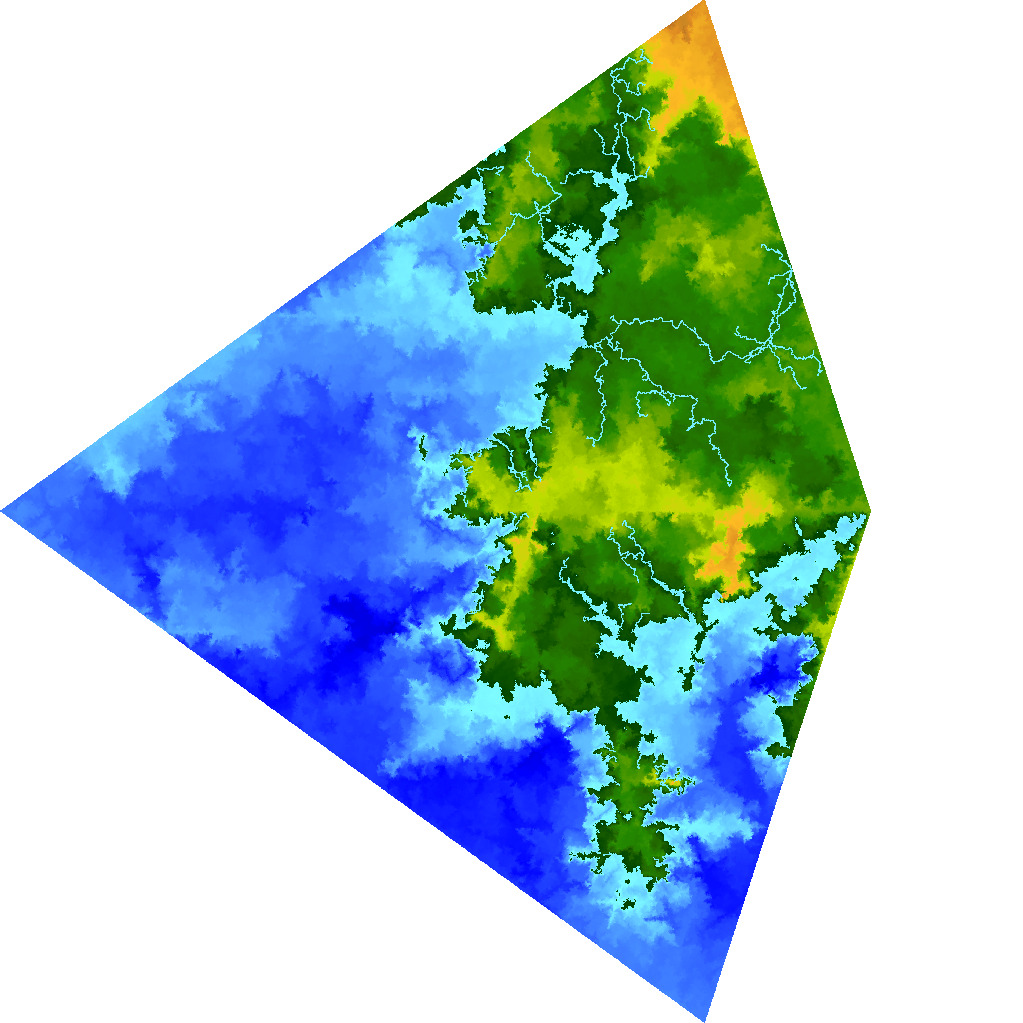}
  \end{center}
\caption{Penrose-tiled map}\label{kite-map}
\end{figure}

To reduce the axis-aligned artefacts, we can use Penrose
tiles~\cite{gardner:1977,penrose:1979,Gardner:1989}, which form
aperiodic tilings.  We use the P2-tiling that uses ``kites and darts'',
both formed by joining two Robinson triangles: isoscleles triangles
where the ratio of the long and short sides is the golden ratio $\phi
= \frac{\sqrt{5}+1}{2}$.  These can be subdivided into smaller
Robinson triangles as shown in Figure~\ref{Robinson}.  The divided
edge is split at a ratio of $\phi$ to 1.  Since the split is uneven,
mirroring matters, so we have indicated the orientation of triangles
with arrows.  Two of the top-type (acute) triangles form a kite shape,
and a map in this shape created by subdividing these is shown in
Figure~\ref{kite-map}.  We have used the same rules for river
placement as for the right-angles triangles, but the altitude of a new
vertex on a split edge without rivers use a weighted average of the
endpoints (with random displacement), so the altitude of the new point
is (statistically) closer to the nearest of its neighbours.  We use

\[V_3.h = (\phi-1)V_1.h+(2-\phi)V_2.h + \delta V_3.s\]

\noindent
where $V_1$ is the vertex closest to the cut point.

\section{Assessment}

The method generates zoomable maps with rivers that look somewhat
realistic, but they suffer from the same problems that maps generated
by recursive subdivision usually do: There are some artefacts due to
the chosen grid, and the rivers do not always flow naturally: They do
not always take the steepest route down, and sometimes there are no
rivers where you would expect rivers to be.  Only altitudes and rivers
are generated, other details such as vegetation and climate are not
handled.

Nevertheless, we believe the maps generated by the method can be used
for gaming -- both computer and table-top games -- possibly with
extensions for creating vegetation and climate details.

The values for the constants $k_1,\ldots\,k_6$ are chosen somewhat
arbitrarily by looking at generated maps and see what ``looks
good''. There are possibilities for varying these and even adding more
``magic'' constants to modify the statistics of the produced maps.

The maps shown in the paper are all made by a straight-forward
implementation of the algorithm in F\# running on Mono, and all use
$1023×1023$ pixels.

Relatively few edges have rivers, so even though handling rivers is
fairly complex, this is done relatively rarely, so rivers add almost
nothing to the running time.  The time to generate unzoomed maps with
or without rivers is about 1.4 seconds.  The $125×$ zoomed map takes
about 1.5 seconds, so the time is also almost independent on the zoom
factor.  All executions are done using single-threaded F\# compiled to
.NET and executed using Mono on an Intel Core i7-10510U processor
running at 1.80GHz.

Using integers instead of floating points for coordinates, seeeds, and
altitudes can speed calculation up, but the reference implementation
uses floats to be close to the description in this paper.  It is also
fairly easy to exploit multiple cores: The first few subdivisions can
make one of the recursive calls run on a different core.

The Penrose-tile variant takes a bit longer to generate, and it can
not (easily) use integer coordinates.  But it generates fewer
grid-aligned artefacts than the regular subdivision method.

\section{Future Work}

\begin{figure}
\setlength{\unitlength}{0.5ex}
\begin{center}
\begin{picture}(50,60)(0,10)

\put(10,20){\line(3,-1){30}}
\put(10,20){\line(2,1){40}}
\put(10,20){\line(2,5){20}}

\put(40,10){\line(1,3){10}}
\put(40,10){\line(-1,6){10}}

\put(30,70){\line(2,-3){20}}

\put(20,45){\circle*{1}}

\qbezier[30](20,45)(30,27.5)(40,10)
\qbezier[30](20,45)(35,42.5)(50,40)

\end{picture}
\end{center}
\caption{Cutting a tetrahedron in two\label{tetracut}}
\end{figure}
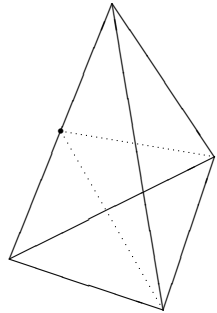

Future work may include extending the method to tetrahedral
subdivision~\cite{Mogensen:2009}.  This generates a planet map by
projecting each map point onto a sphere contained in an irregular
tetrahedron, and then recursively subdividing this while discarding
tetrahedra that do not contain the point.  Figure~\ref{tetracut}
illustrates tetrahedral subdivision.

While tetrahedral subdivision resembles triangle subdivision, adding
rivers is not trivial: With triangles, rivers can float from edges to
edges, as all edges will be part of the final map.  With tetrahedra,
an edge may or may not intersect the surface of the sphere (and may do
so twice).  It may work to say that a river can only be generated on
an edge that intersects the sphere exactly one, and when subdividing
the edge, the river is put on the half edge that intersects the
sphere, but that is unlikely to give natural-looking river flows.
Additionaly, an edge can be shared by more than two tetrahedra, so
placing a river there is likely to give three-way or four-way river
joins.  It may be better to place rivers on faces instead of edges, as
a face is shared only by two tetrahedra.  Again, we can generate a
river on a face only if it intersects the sphere, and when subdividing
a face (which happens when a tetrahedron is cut in two), the river can
be placed on a sub-face that intersects the sphere.  There may be a
choice, which can be resolved similar to how it is done with triangle
subdivision.  A problem is that rivers on faces can not affect
altitudes on the map, as the altitude on the new point can only depend
on attributes of the edge on which it is positioned and the end points
of this edge.  So it may be impossible to make rivers flow downwards
only.

\bibliography{rivers2.bbl}

\end{document}